\documentclass[pre,twocolumn,showpacs,floatfix]{revtex4}

\usepackage{graphicx,rotating}
\usepackage{bm,amssymb}

\newcommand{\rem}[1]{}

\newcommand{\eps}{\varepsilon}
\newcommand{\mx}{{\text{max}}}
\newcommand{\tot}{{\text{tot}}}
\newcommand{\step}{{\text{st}}}

\newcommand{\cN}{{\cal N}}

\begin{document}

\title{Low rank perturbations and the spectral statistics of 
   pseudointegrable billiards}

\author{Thomas Gorin}
\affiliation{Theoretische Quantendynamik, Universit\"at Freiburg, 
   Hermann-Herder-Str 3, D-79104 Freiburg}
\email{Thomas.Gorin@physik.uni-freiburg.de}
\author{Jan Wiersig}
\affiliation{Institut f\"ur Theoretische Physik, Universit\"at Bremen, 
  Kufsteiner Str., D-28359 Bremen, Germany}
\date{\today}

\begin{abstract} 
We present an efficient method to solve Schr\"odinger's equation for
perturbations of low rank. In particular, the method allows to calculate the 
level counting function with very little numerical effort. To illustrate the 
power of the method, we calculate the number variance for two pseudointegrable 
quantum billiards: the barrier billiard and the right triangle billiard 
(smallest angle $\pi/5$). In this way, we obtain precise estimates for the 
level compressibility in the semiclassical (high energy) limit. In both cases, 
our results confirm recent theoretical predictions, based on periodic orbit
summation.
\end{abstract}


\maketitle

Consider the stationary Schr\"odinger equation for a quantum system, which can
be described by a Hamiltonian $H= H_0 + W$  with the following properties: 
The operators $H$ and $H_0$ have discrete spectra, the eigenbasis of 
$H_0$ is known, and the perturbation $W$ is non-negative (or non-positive) and
of low rank. Then, our method allows to evaluate the level counting function 
at arbitrary energies, by solving an eigenvalue problem of the dimension which 
is equal to the rank of the perturbation. At first sight, the requirements 
seem rather restrictive, but in fact, there are a number of problems 
which can comply with them. For example, one may think of a few particle 
system with some kind of short range residual interaction.

To illustrate how the method works, we choose two examples of a different 
type, the so called ``barrier billiard''~\cite{Hannay90}, and the right 
triangle billiard~\cite{Gut1+2} with smallest angle $\pi/5$. Both examples are 
twodimensional, pseudointegrable polygon 
billiards~\cite{RichensBerry81,Gut1+2}. While pseudointegrable systems have 
enough constants of motion to assure local integrability, singularities in the 
Hamiltonian flow allow invariant surfaces with genus larger than one.

In the spirit of quantum-classical correspondence, there have been numerous
efforts to study the implications of pseudointegrability on the quantum
spectrum~\cite{RichensBerry81,GreJai98,BGS99,CasPro99b,Gorin01,Wiersig02}. In 
Refs.~\cite{BGS99,BGS01b} it is conjectured that the statistical properties of 
pseudointegrable billiards is intermediate between Poisson statistics, 
typically related to integrable systems~\cite{BerryTabor77a}, and the 
eigenvalue statistics of the Gaussian orthogonal ensemble~\cite{Mehta91},
related to fully chaotic (time reversal invariant) systems. The so called
``intermediate statistics'' has also been found in disordered, mesoscopic 
systems at the metal-insulator transition~\cite{SSSLS93}, for systems with 
interacting electrons~\cite{WWP99}, and for incommensurate double-walled 
carbon nanotubes~\cite{AKWC02}.

A suitable measure for intermediate statistics is the level compressibility 
$\chi= \lim_{L\to\infty} \Sigma^2(L)/L$, where $\Sigma^2(L)$ is the number 
variance~\cite{Mehta91} for energy intervals of length $L$ (measured in units 
of the average level spacing). Note that $\chi$ coincides with the value of 
the spectral two-point formfactor in the limit of small times. Recently, 
analytical results for $\chi$ became available for a certain class of
right triangle billiards (including our example)~\cite{BGS01}, as well as 
for the barrier billiard~\cite{Wiersig02}. However, the convergence to the 
semiclassical limit is so slow, that numerical studies could not confirm 
those results with the desirable clarity. For example in the case of 
the barrier billiard one obtains $\chi\approx 0.34$ in the region between 
level number $400\,000$ and $420\,000$, while the semiclassical prediction is 
$\chi = 1/2$~\cite{Wiersig02}. 

As we will show below, our method is almost 
ideally suited for the calculation of the number variance for large values 
for $L$. Therefore, we are able to calculate the level compressibility at 
much higher energies and with better statistics. In the case of the barrier 
billiard, for example, we calculate number variances in the region of absolute 
level number $\cN > 1.6*10^7$ within an energy range which contains about
$10^5$ levels. \\

Here we can only sketch the general method. A detailed presentation will be
published elsewhere, and an early implementation can be found in
Ref.~\cite{Gorin01}. Assume, that the Hamiltonian $H$ can be approximated by
a projection on an appropriately chosen $N$-dimensional Hilbertspace, 
$N < \infty$. Furthermore, assume that its matrix representation is of the 
following form:
\begin{equation}
H= H_0 + \eta\; V\, V^\dagger \; .
\end{equation}
Here $V$ is a $N$$\times$$M$-matrix with mutually orthogonal column vectors,
$M < N$, and $\eta$ is a positive parameter. Such a representation can be 
obtained, for instance, by diagonalizing a non-negative perturbation of small
rank $M$. Then, in general, the spectrum of $H$ consists of a trivial 
component $S_0$, contained in the spectrum of $H_0$, and a non-trivial 
component $S_1$, which is disjoint. The eigenvalues in $S_1$ are roots of the 
secular equation
\begin{equation}
\det[1-\eta K(E)] = 0\; , \quad K(E)=  V^\dagger \; \frac{1}{E-H_0}\; V \; ,
\label{seceq}\end{equation}
where the matrix $K(E)$ is of dimension $M$~\cite{Gorin01}.
Eq.~(\ref{seceq}) can be considered as an eigenvalue equation
$K(E)\, \vec x = \delta\, \vec x$, where one of the eigenvalues must be 
equal to $\eta^{-1}$. 

Differentiating $K(E)$ with respect to the energy gives a negative definite
matrix. This shows that the eigenvalues of $K(E)$ are
monotonously decreasing with energy. The eigenvalues of $H_0$ coincide with 
the positions of the poles of $K(E)$. Therefore, each time the energy moves
across a pole, a new eigenvalue of $K(E)$ appears at $+\infty$. The eigenvalue
decreases with energy, until it reaches the value $1/\eta$. At this point the
secular equation~(\ref{seceq}) has a root. Beyond this point, the eigenvalue
continues to decrease, until it disappears at $-\infty$. This behavior gives 
rise to the following {\em sum rule}: Let $n_0(n_1)$ denote the number of 
eigenvalues of $K(E)$, larger than $1/\eta$,  at $E= E_0(E_1)$, and let $N_p$ 
denote the number of poles and $N_r$ the number of roots of $\det[K(E)]$ in 
the interval $(E_0,E_1)$. Then it holds:
\begin{equation}
n_0 + N_p - N_r = n_1 \; .
\label{sumrule}\end{equation}
At the one hand, this relation can be used to bracket the eigenvalues in $S_1$,
before using a standard root searching algorithm to obtain the desired 
accuracy. In this case, Eq.~(\ref{sumrule}) assures, that no eigenvalues
are overlooked. On the other hand, in order to study the long range 
correlations, it can be sufficient to calculate the level counting
function at the points of an equi-distant grid (in units of the average level 
spacing) grid in the energy region of interest. If $L_\step$ denotes the
stepsize, one can obtain the number variance at integer multiples of 
$L_\step$ from this data. \\

To calculate the spectrum of the barrier billiard~\cite{Wiersig02}, we choose
as $H_0$ the Hamiltonian of a rectangle billiard with sides of length $a$ and
$b$. With the origin of a Cartesian coordinate system fixed at one of the 
corners, the normalized eigenfunctions are
\begin{equation}
\Psi_{mn}(x,y) = 2(ab)^{-1/2}\; \sin(\pi m x/a) \; \sin(\pi n y/b)\; ,
\label{B_Psis}\end{equation}
while $\eps_{mn}= [(m/a)^2 + (n/b)^2]\, \pi^2/2$ are the corresponding 
eigenvalues. We use units, in which the mass and Planck's constant $\hbar$ are 
both equal to one. The perturbation consists of an additional boundary segment 
inside the billiard, connecting the points $(a_0,0)$ and $(a_0,c)$. It is 
modeled by a potential well with delta shaped profile:
\begin{equation}
H = H_0 + \eta W \; , \quad   W = (a/2)\; \delta(x-a_0) \; \theta(c-y) ,
\label{Ham}\end{equation}
where $\delta(x)$ is the usual $\delta$-function, and $\theta(y)$ is the
unit step function. The prefactor $a/2$ is included for convenience. As $\eta$ 
increases from 0 to $\infty$, the spectrum of $H$ changes from the spectrum of 
the rectangle billiard to the spectrum of the barrier billiard.

In what follows we set $a_0 = a/2 ,\, c= b/2$, where $a= 2\pi^{3/2}/3$ and 
$b= 6/\pi^{1/2}$. In this way, we obtain the same spectrum as in 
Ref.~\cite{Wiersig02}. In this case the trivial component $S_0$ of the 
spectrum corresponds to states $\Psi_{mn}$ with a 
node line at $x= a/2$. In order to obtain the non-trivial component alone, we 
require the eigenstates to be reflection symmetric with respect to that line.
In other words, we consider the spectrum of a new rectangle billiard with 
sides $a/2$ and $b$, which has Dirichlet boundaries everywhere, except for the 
boundary segment between the points $(a/2,b/2)$ and $(a/2,b)$ where it 
has a von Neumann boundary.

To obtain the decomposition $W= V\, V^T$, we calculate all 
eigenvectors of $W$ which correspond to nonzero eigenvalues. This is rather 
simple because $W$ is separable in the $x$- and $y$-modes of the 
eigenfunctions of $H_0$:
\begin{eqnarray}
&& V_{mn}^{(\alpha)} = s(m)\; \tilde A_n^{(\alpha)} \; \quad 
s(m)= \sin(\pi m/2) \\
&&\tilde A_{2n}^{(\alpha)} = \frac{\delta_{n\alpha}}{\sqrt{2}}\; \quad
\tilde A_{2n-1}^{(\alpha)} = \frac{\sqrt{2}\; \alpha}{\pi}\;
   \frac{(-)^{\alpha+n}}{\alpha^2 - (n-1/2)^2} \; . 
\end{eqnarray}
For the matrix elements of $K(E)$ we get:
\begin{equation}
K_{\alpha\beta}(E) = \sum_{mn} \frac{V_{mn}^{(\alpha)} \; V_{mn}^{(\beta)}}
   {E-\eps_{mn}} = \sum_{mn} \tilde A_n^{(\alpha)}\; \tilde A_n^{(\beta)}\;
   \frac{s(m)^2}{E-\eps_{mn}} \; .
\end{equation}
In the limit $m\to\infty$, we can evaluate the sum over $m$ analytically:
\begin{equation}
K_{\alpha\beta}(E) = \frac{a^2}{2\pi}\sum_n \tilde A_n^{(\alpha)}\, 
   \tilde A_n^{(\beta)}\, G_n(E) \; ,
\label{BS_Gn}\end{equation}
with $G_n(E)= - \tan(\pi z_n/2)/z_n$. Here, $z_n$ is the effective 
quantum number for $H_0$ at given energy $E$, {\it i.e.} 
$(z_n\pi/a)^2 + (n\pi/b)^2 = 2E$. Note that $G_n(E)$ remains real, even for
imaginary $z_n$. Introducing the orthogonal matrix
$A_{nm} = \sqrt{2}\, \tilde A_{2m-1}^{(n)}$ we may write:
\begin{equation}
K(E)= \frac{a^2}{4\pi}(A\; G^{\rm odd}\; A^T + G^{\rm even}) \; ,
\label{Kmat}\end{equation}
with $G^{\rm odd} = {\rm diag}[G_{2n-1}(E)]$, and 
$G^{\rm even} = {\rm diag}[G_{2n}(E)]$. Multiplying Eq.~(\ref{Kmat}) from left 
and/or right by $A^T$ and/or $A$, one can construct additional variants: 
$L= A^T\, K$, $K'= A^T\, K\, A$, and $M= K\, A$. Any one can
be used in the secular equation~(\ref{seceq}) to find the eigenvalues of the
barrier billiard. However, the matrices 
$K(E)$ and $K'(E)$ are real and symmetric, and hence easy to diagonalize. For 
the analysis below, we use $K'(E)$, because it turned out that the numerical 
procedure converges faster in this case. In Ref.~\cite{Wiersig02} the 
non-symmetric matrix $M(E)$ has been used, instead. \\

In what follows, we aim at a precise, numerical estimate for the
level compressibility, which can be compared to the analytical results 
obtained in \cite{BGS01b,Wiersig02}. To this end, we compute the level counting
function on a finite grid $\cN_0, L_\tot, L_\step$ of consecutive intervals of 
length $L_\step$, starting at $\cN_0$ and ending at $\cN_0 + L_\tot$. The 
grid is defined on the unfolded energy axis, where the average level spacing 
is equal to one. It is mapped onto the physical energy axis, by inverting the 
Weyl law~\cite{BraBha97}:
\begin{equation}
\cN(E) = \frac{ab}{4\pi}\; E - \frac{a+b}{4\pi}\; \sqrt{2E} + \frac{1}{16} \; .
\end{equation}
We count the number of levels in each interval, using the sum rule, 
Eq.~(\ref{sumrule}). The data sets, listed in Tab.~\ref{data-H}, are produced
in this way. Finally, they are used, to calculate the number variance 
$\Sigma^2(L)$ at integer multiples of $L_\step = 10$. To get an idea of the 
efficiency of the method, note that it took about twenty days on a PC running 
at 1.7 GHz, to produce the data set (D). The dimension of the matrices, to be 
diagonalized, was approximately 3700.

\begin{table}
\begin{tabular}{|rcccc|}
\hline
          & (A)      & (B)        & (C)      & (D)        \\
\hline
$\cN_0=$  & $8*10^5$ & $2*10^6$   & $8*10^6$ & $1.6*10^7$ \\
$L_\tot=$ & $10^5$   & $1.4*10^5$ & $10^5$   & $10^5$     \\
\hline
\end{tabular}
\caption{Data sets for the barrier billiard.}
\label{data-H}\end{table}

According to Ref.~\cite{Berry85}, the number variance $\Sigma^2(L)$ saturates 
at $L \approx L_\mx$, which is related to the inverse period of the shortest 
(non-diffractive) periodic orbit. For our barrier billiard
\begin{equation}
L_\mx = \sqrt{2\pi\, {\cal N}_0\; b/a} = \sqrt{18\; {\cal N}_0/\pi}\; .
\end{equation}
At large $L\gg 1$, the ratio $\Sigma^2(L)/L$ is constant at first, but it
starts to decrease, when $L$ approaches $L_\mx$. Therefore, we may expect that 
a plot: $\Sigma^2(L)/L$ versus $L/L_\mx$ depends only weakly on $L_\mx$,
{\it i.e.} the energy range $\cN_0$. We 
call this function the {\em scaled number variance}:
\begin{equation}
g(x)= \Sigma^2(L_\mx\, x)/(L_\mx\, x) \; .
\end{equation} 
In theory it is expected that in a given energy region, the ratio 
$\Sigma^2(L)/L$ approaches its limit value quite quickly; for instance,
$L \approx 10$ seems to be sufficient, to obtain accurate results. The 
semiclassical limit $\cN_0\to\infty$ is the really difficult one. In this 
limit the level compressibility must be estimated by extrapolation: 
$\chi = \lim_{x\to 0} g(x)$.

\begin{figure}
\setlength{\unitlength}{1mm}
\begin{picture}(86,62)
\put(4,4){\includegraphics[scale=0.66]{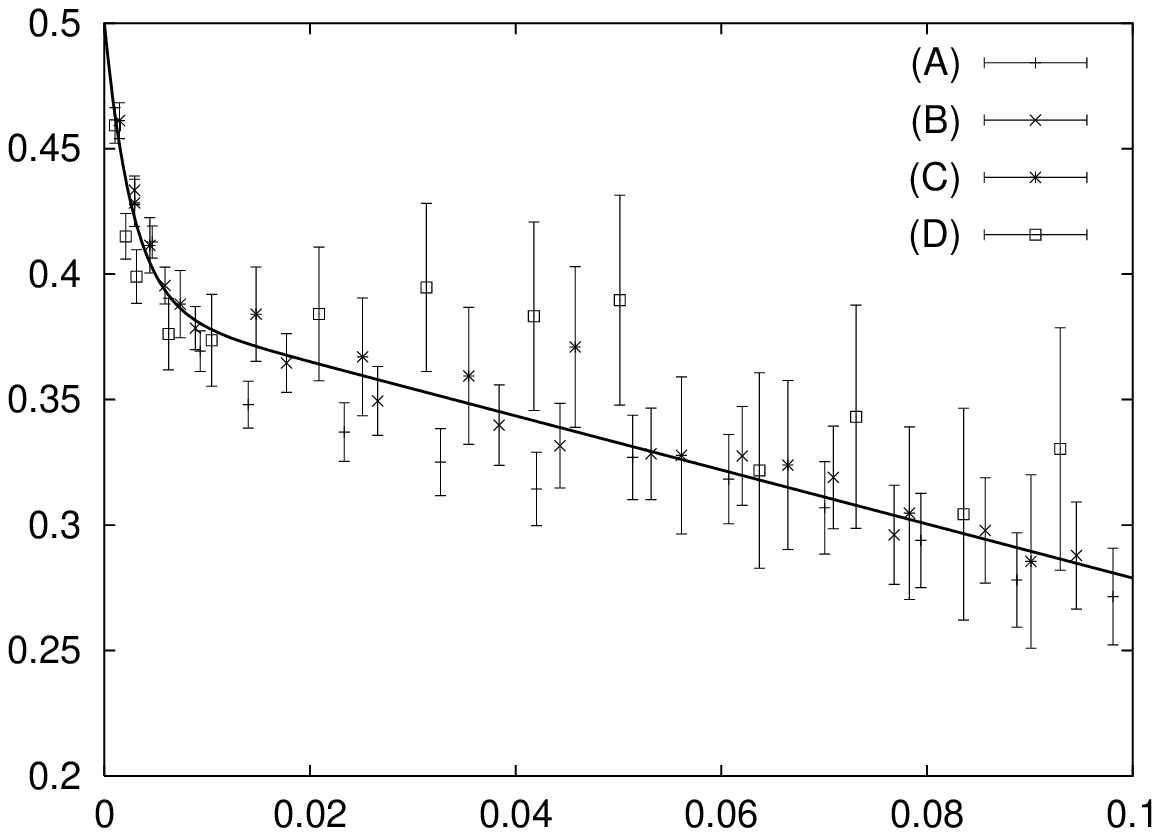}}
\put(49,2){\makebox(0,0){$L/L_\mx$}}
\put(2,35){\makebox(0,0){\begin{sideways} $\Sigma^2(L)/L$\end{sideways}}}
\end{picture}
\caption{The scaled number variance $\Sigma^2(L)/L$ versus $x= L/L_\mx$, for 
the four data sets given in Tab.~\ref{data-H}. For clarity, only some of the 
data points are plotted. The thick solid line gives the fit function $g(x)$ 
with parameters as given in Eq.~(\ref{fit-1}).}
\label{F_sig-2d}
\end{figure}

Fig.~\ref{F_sig-2d} shows the scaled number variance 
for the data sets given in Tab.~\ref{data-H}. For the error bars, we
estimated that the relative error is approximately equal to 
$1.6\, \sqrt{L/L_\tot}$. In order to check this, we computed the 
variance of the distribution of different partial averages, fitting them
with a normal distribution. Note that $L_\tot/L$ gives the number of 
independent level counts, in the energy range considered. Fig.~\ref{F_sig-2d} 
explains the discrepancy between the numerical estimate for the level 
compressibility obtained in Ref.~\cite{Wiersig02}, and the theoretical 
expectation. In the region $x > 0.01$ the scaled number variance looks 
perfectly linear. Hence, in the absence of data points with $x < 0.01$, one is 
lead to assume that the linear behavior continues down to the 
point $x=0$. However, for $x < 0.01$, the slope changes drastically, and 
a second linear regime appears. Then indeed, the scaled number variance $g(x)$ 
approaches the theoretical prediction $\chi= 1/2$ as $x\to 0$. In order to 
put our findings on a quantitative basis, consider the following 
phenomenological parametrization:
\begin{equation}
g(x)= a_2 - a_1 x + (a_0-a_2) \exp(-a_3 x) \; .
\label{fitfun}\end{equation}
Its form is such that $a_0= g(0)$ gives the best estimate for the level 
compressibility in the semiclassical limit. At the same time, both linear 
regimes are reproduced. By consequence, $a_2$ would be the estimate for the 
level compressibility, in the absence of data points with $x< 0.01$. For the 
fit, all data sets in Tab.~\ref{data-H} are taken into account, excluding only
those data points, for which $x > 0.2$. Beyond this point (which is outside 
the interval shown in Fig.~\ref{F_sig-2d}) the parametrization~(\ref{fitfun})
breaks down. Using the nonlinear least squares Marquardt-Levenberg 
algorithm~\cite{gnu99}, we obtain the following estimates for the 
fit-parameters:
\begin{eqnarray}
a_0& = &0.5008(91) \qquad a_1 = 1.077(18)  \nonumber\\
a_2& = &0.3866(17) \qquad a_3 = 364(39)
%
%
\label{fit-1}\end{eqnarray}
with a reduced chi-square value of $\chi_{\rm fit}^2/f \approx 0.54$ (the
number of degrees of freedom is $f = 305$).

With $a_0$, we have finally obtained a precise numerical estimate for the level
compressibility in the semiclassical limit. It agrees within a relative error
of roughly two percents with the theoretical result $\chi = 1/2$. In 
addition, the moderate value for $\chi_{\rm fit}^2$ is quite remarkable. It 
gives some support to the assumption that within the statistical error, 
the scaled number variance is independent of the energy region 
(as long as $x\lesssim 0.2$). \\

\begin{table}
\begin{tabular}{|rccc|}
\hline
          & (T1)     & (T2)   & (T3)     \\
\hline
$\cN_0=$  & $4*10^5$ & $10^6$ & $4*10^6$ \\
$L_\tot=$ & $4*10^4$ & $10^5$ & $10^5$   \\
\hline
\end{tabular}
\caption{Data sets for the right triangle billiard ($\alpha= \pi/5$).}
\label{data-Tri}\end{table}

\begin{figure}
\setlength{\unitlength}{1mm}
\begin{picture}(86,62)
\put(4,4){\includegraphics[scale=0.66]{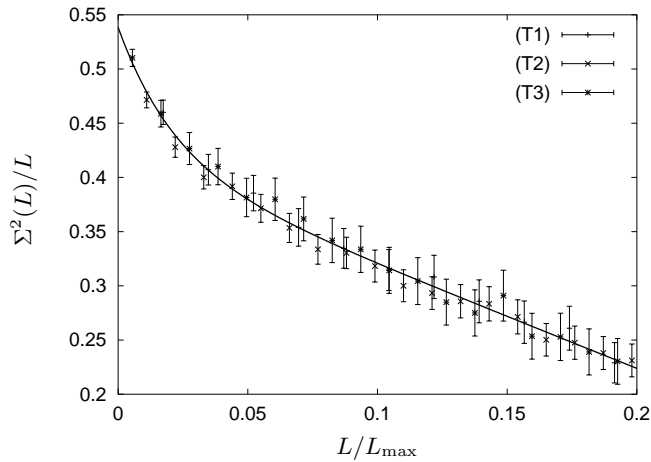}}
\put(49,2){\makebox(0,0){$L/L_\mx$}}
\put(2,35){\makebox(0,0){\begin{sideways} $\Sigma^2(L)/L$\end{sideways}}}
\end{picture}
\caption{The scaled number variance for
the data sets in Tab.~\ref{data-Tri}. Only a subset of the 
data points is plotted. The solid line gives the fit $g(x)$ 
with parameters as given in Eq.~(\ref{fit-tri1}).}
\label{F_trisig-2d}
\end{figure}

In what follows, we repeat the numerical analysis for the 
$\pi/5$-right triangle billiard. In this case the length scale $L_\mx$, for 
the saturation of the number variance, is about $2.6$ times smaller than for 
the barrier billiard, {\it i.e.}
\begin{equation}
L_\mx = \sqrt{2\pi\, \cN_0/(8\, \sin 2\alpha)} \qquad
\alpha= \pi/5 \; .
\end{equation}
We calculated three data sets in different energy regions, as listed
in Tab.~\ref{data-Tri}. The stepsize was again $L_\step= 10$. With all data 
sets, we fit the scaled number variance $g(x)$ using the 
parametrization~(\ref{fitfun}), excluding as before all data points with 
$x > 0.2$. The resulting estimates are:
\begin{eqnarray}
a_0& = &0.5385(57) \qquad a_1 = 0.958(37) \nonumber\\
a_2& = &0.4154(54) \qquad a_3 = 46.2(5.4)
%
\label{fit-tri1}\end{eqnarray}
with a reduced chi-square value of $\chi_{\rm fit}^2/f \approx 0.20$ 
($f = 61$). The numerical results for $g(x)$, and the fit with 
Eq.~(\ref{fitfun}), are plotted in Fig.~\ref{F_trisig-2d}.
For the error bars, we have checked, that the same approximation holds as in
the case of the barrier billiard. The absolute errors are smaller here, 
because the data sets are taken at lower energy ranges. As in 
Fig.~\ref{F_sig-2d}, we find two different linear regimes, with a transition
which occurs somewhat earlier, at $x\approx 0.04$. The overall change in the 
graph is less dramatic because the slope at $x < 0.04$ is much smaller in 
magnitude. Though the agreement of the extrapolated value for $g(0)$ with the 
theoretical expectation $\chi = 5/9$~\cite{BGS01} is not perfect, our 
estimate is quite close to it. Note that the error estimates for the 
parameters $a_0,\ldots,a_3$ are also based on statistical data. In particular 
in the case of $a_0$, there are only relatively few relevant data points. 
This leads to rather large uncertainties. In general the data points 
fluctuate much less than in the case of the barrier billiard, as can be read
off from the smaller value for $\chi_{\rm fit}^2/f$. \\

We advanced the technique, recently proposed in~\cite{Gorin01}, to solve 
the Schr\"odinger equation for perturbations of low rank. We realized
that due to a special {\em sum rule}, one can obtain the level counting
function with very little numerical effort. Systems involving short range
interactions may be good candidates for future applications. In this letter, 
we considered two pseudointegrable billiards: the barrier billiard and
the $\pi/5$-right triangle billiard. In spite of their apparent simplicity,
the spectral statistics, a corner stone in quantum-classical correspondence,
is only hardly understood. We performed extensive numerical calculations for
the number variance $\Sigma^2(L)\, , L\gg 1$ in energy regions up to 
$\cN \ge 1.6*10^7$ (for the barrier billiard) and $\cN \ge 4.*10^6$ (for the
right triangle billiard). With the help of the {\em scaled number 
variance}, we obtained very precise estimates for the level compressibility,
which largely confirm previously obtained analytical results.
%

T.G. thanks L.~Kaplan and T.~Papenbrock for stimulating discussions
at the Centro Internacional de Ciencas in Cuernavaca, Mexico, and 
the EU Human Potential Program contract HPRN-CT-2000-00156 for financial 
support.

\bibliographystyle{prsty}
\bibliography{../bib/fg4,../bib/extern}

\end{document}